# Ear-clipping Based Algorithms of Generating High-quality Polygon Triangulation

Gang Mei[1], John C.Tipper[1] and Nengxiong Xu[2]

**Abstract** A basic and an improved ear-clipping based algorithm for triangulating simple polygons and polygons with holes are presented. In the basic version, the ear with smallest interior angle is always selected to be cut in order to create fewer sliver triangles. To reduce sliver triangles in further, a bound of angle is set to determine whether a newly formed triangle has sharp angles, and edge swapping is accepted when the triangle is sharp. To apply the two algorithms on polygons with holes, 'Bridge' edges are created to transform a polygon with holes to a degenerate polygon which can be triangulated by the two algorithms. Applications show that the basic algorithm can avoid creating sliver triangles and obtain better triangulations than the traditional ear-clipping algorithm, and the improved algorithm can in further reduce sliver triangles effectively. Both of the algorithms run in $O(n^2)$ time and $O(n)$ space.

## 1 Introduction

Polygons are very convenient for representing real objects. However, in some cases polygons are too complex. In order to implement polygons faster and easier in applications, usually polygons need to be decomposed into simpler components such as triangles [2,3], trapezoids [12] or even sub-polygons [5].

In computational geometry, polygon triangulation is the decomposition of a polygonal area into a set of triangles [1,7], or in other words, to create a set of triangles without non-intersecting interiors whose union is the original polygon.

Gang Mei, John C.Tipper (✉)
Institut für Geowissenschaften – Geologie, Albert-Ludwigs-Universität Freiburg, Albertstr. 23B, D-79104, Freiburg im Breisgau, Germany
e-mail: {gang.mei, john.tipper}@geologie.uni-freiburg.de

Nengxiong Xu (✉)
School of Engineering and Technology, China University of Geosciences, Beijing,100083, China
e-mail: xunengxiong@yahoo.com.cn

One way to triangulate a simple polygon is based on a fact that any simple polygon with at least 4 vertices without holes has at least two 'ears', which are triangles with two sides being the edges of the polygon and the third one completely inside it. This fact was proved by Meisters [10]. Meisters proposed a recursively algorithm that consists of searching an ear and cutting it off from current polygon. Removing an ear results in forming a new polygon that still meets the 'two ears' condition and repetitions can be done until there is only one triangle left.

The directly implemented ear clipping method runs in $O(n^3)$ time, with $O(n)$ time spent on checking whether a triangle newly constructed is valid. But in 1990 an efficient technique named 'prune-and-search' brought the time complexity from $O(n^3)$ down to $O(n^2)$ [4]. Also, a simple reorganization of Meisters's algorithm leads to run ear clipping algorithm in time complexity of $O(n^2)$ [11].

There are other triangulation algorithms proposed based on ear clipping, such as Kong, Everett and Toussaint algorithm [9], which adopts the Graham scan to select ears. This algorithm is sensitive to the shape of the polygon and runs in $O(n(r+1))$ time, where $r$ denotes the number of reflex vertices of the polygon.

A triangle with sharp angle that is also called *silver triangle* is not allowed for its poor shape quality. Based on Rourke's algorithm, Sloan [6] also developed an $O(n^2)$ ear-clipping algorithm. After completely obtaining the triangulation of a polygon, optimization by swapping diagonals is accepted to avoid sliver triangles. Held [8] developed a polygon triangulation package FIST, which is based on ear clipping and can be applied to deal with complex polygons.

The motivation of this paper is to design a new algorithm based on ear clipping to generate high-quality triangulations with fewer sliver triangles. To achieve this, when locate an ear tip, the one with smallest interior angle is always selected and removed. If the ear tip with smallest angle is not chosen firstly, it will be divided into at least two much smaller ones, and several sharp triangles will be formed. To reduce *sliver triangles* in further, edge swapping is adopt during cutting ears rather than after generating polygon triangulation as that done in Sloan's code [6].

In our algorithm, firstly all vertices of a polygon are determined to be convex or reflex according to their interior angles. Secondly, all ear tips will be found by temporarily forming a triangle with each vertex $v_i$ and its two adjacent vertices $v_{i-1}$ and $v_{i+1}$, and then testing whether all reflex vertices are in the triangle $\triangle(v_{i-1}, v_i, v_{i+1})$. If inside, $v_i$ is an ear tip; Otherwise it's not. After identifying all ear tips, the ear tip $v_i$ with smallest angle is selected, the ear consisted by three vertices ($v_{i-1}$, $v_i$, $v_{i+1}$) is removed and a triangle $\triangle(v_{i-1}, v_i, v_{i+1})$ can be formed. After removing, ear tip status must be updated for vertices $v_{i-1}$ and $v_{i+1}$. And repetitions can be done until there is only one triangle left.

This paper is organized as follows. Sect.2 describes the basic triangulation algorithm for simple polygons and its improved version with edge swapping. Both of the proposed algorithms are extended to the polygons with holes. In Sect.3, tests are made, qualities of resulting triangulations are compared and complexity is analyzed. Finally, Sect.4 concludes the work.

## 2 The Proposed Algorithms

### 2.1 Basic Algorithm

The ear clipping triangulation algorithm consists of searching an ear and then cutting it off from current polygon. The original version of Meisters's ear clipping algorithm runs in $O(n^3)$ time, with $O(n)$ time spent on checking whether a newly formed triangle is valid. Rourke [11] simply modified and reorganized Meisters's algorithm and made the new version of ear clipping algorithm run in $O(n^2)$ time. The algorithms proposed in this paper are based on Rourke's algorithm.

Given a simple polygon $P$ with $n$ vertices $V(v_0, v_1, \ldots, v_{n-1})$ orientated in countclockwise(CCW), the basic algorithm proposed in the paper to triangulate simple polygon $P$ is designed into four steps. Pseudocode is listed as Algorithm 1.

Step 1: Compute the interior angles on each vertex of $P$. If the interior angle on a vertex is less than 180°, the vertex is convex; Otherwise, reflex.

Step 2: Find out all ear tips of $P$, and initiate the ear tip status for each vertex according to the following Condition 1 [8, 11, 14].

**Condition 1** Three consecutive vertices $v_{i-1}$, $v_i$, $v_{i+1}$ of $P$ do form an ear if

1. $v_i$ is a convex vertex;

2. the closure $C(v_{i-1}, v_i, v_{i+1})$ of the triangle $\triangle(v_{i-1}, v_i, v_{i+1})$ does not contain any reflex vertex of $P$ (except possibly $v_{i-1}, v_{i+1}$). The closure of a triangle is formed by the union of its interior and its three boundary edges.

Step 3: Select the ear tip $v_i$ which has smallest angle to create a triangle $\triangle(v_{i-1}, v_i, v_{i+1})$, and then delete the ear tip $v_i$, update the connection relationship, angle and ear tip status for $v_{i-1}$ and $v_{i+1}$.

Step 4: Repeat Step 3 until $(n-2)$ triangles are constructed.

Both Step 1 and Step 2 are to initiate status for the original polygon, while Step 3 and Step 4 are the repetition of cutting and updating.

---

**Algorithm 1** Triangulate Simple Polygon ($P$)

---

*Input*: A simple polygon $P$ with $n$ vertices $V(v_0, v_1, \ldots, v_{n-1})$
*Output*: A triangulation $T$ with $n-2$ triangles
1: Compute interior angles of each vertex in $P$.
2: Indentify each vertex whether it is an ear tip or not.
3: **while** number of triangles in $T < n-2$ **do**
4:     Find the ear tip $v_i$ which has the smallest interior angle.
5:     Construct a triangle $\triangle(v_{i-1}, v_i, v_{i+1})$ and add it onto $T$.
6:     Let $v_i$ be no longer an ear tip.
7:     Update connection relationship of $v_{i-1}$ and $v_i$, $v_i$ and $v_{i+1}$, $v_{i-1}$ and $v_{i+1}$.
8:     Compute the interior angles of $v_{i-1}$ and $v_{i+1}$.
9:     Refresh the ear tip status of $v_{i-1}$ and $v_{i+1}$.
10: **end while**

## *2.2. Improved Algorithm*

The basic algorithm tries to avoid creating sliver triangles. However, in some situations, sliver triangles still appear in triangulations. Thus, edge swapping is accepted to avoid sliver triangles. First of all, a bound for limiting interior angle is set. And then, after the ear tip with smallest interior angle is selected to form a new triangle, the new triangle will be checked whether it is sharp: if one of its three angles is smaller than the bound, the triangle is sharp and needs to swap edge with one of its neighbor. Noticeably, the new triangle may have no neighbor triangle because the triangle itself is created earlier than its neighbors.

If a new triangle needs to optimize, firstly find out its biggest interior angle and its opposite edge (the longest edge), and then search between all the generated triangles to see whether there exists a triangle that shares the longest edge with the new created triangle.

If there is one, the two triangles can form a quadrilateral. And then swapping the diagonal of the quadrilateral to see whether the minimum angle of the original pair of triangles is smaller than the minimum one of the new pair of triangles after swapping, if does, which means the new pair of triangles has better quality than the original one, swapping needs to be done; if not, the original triangles must be kept without swapping (Fig.1).

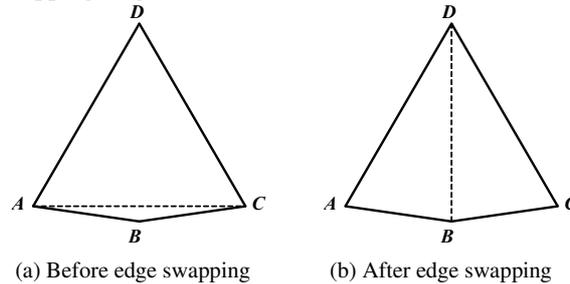

(a) Before edge swapping     (b) After edge swapping

**Fig. 1** Edge swapping

In the procedure of combining the basic algorithm and the edge-swapping optimization, edge swapping just has relationship with the pair of triangles which to be optimized. It does not affect the later updating operations.

## *2.3. Extension to Polygon with Holes*

The basic and improved algorithms have been described above. However, the two algorithms are only valid for simple polygons and can not be directly applied on polygons with holes. Thus, a pre-processing of creating 'Bridge' edges must be done to transform the polygon with holes into a single polygon.

Creating 'Bridge' edges is widely used to divide a general closed domain into several simply connected, convex sub domains, such as generating Delaunay triangulations or Voronoi diagram in multi-domain polygons, see Tipper [13].

Different from dividing a closed domain, Held [8] adopts 'Bridge' edges to transform a multiply-connected polygonal area into a single polygon. The resulting polygon is not a simple polygon since each 'Bridge' edge appears twice with opposite orientations. These polygons are defined as 'degenerate' polygons by Held. In this paper, creating 'Bridge' edges is also accepted. The algorithm of triangulating polygon with holes can be divided into two stages.

Stage 1: Create several 'Bridge' edges to transform a polygon with holes to a degenerate polygon without holes.

Stage 2: Triangulate the resulting polygon in Step 1 by ear clipping.

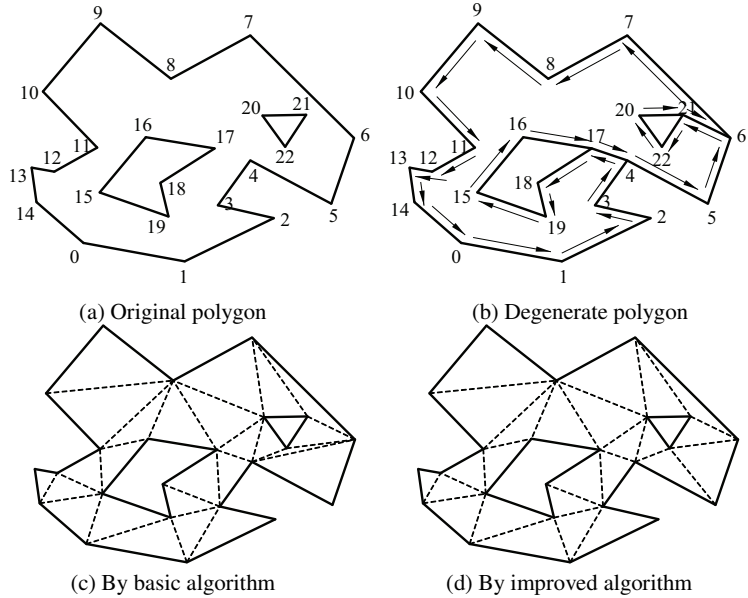

(a) Original polygon　　　　　(b) Degenerate polygon

(c) By basic algorithm　　　　(d) By improved algorithm

**Fig. 2** Transformation of a simple polygon with two holes and its triangulations

Supposing there has a polygon $P$ with $m$ vertices PV($pv_0$, $pv_1$,…,$pv_{m-1}$) which orientate in count-clockwise and a hole $H$ with $n$ vertices HV($hv_0$, $hv_1$,…,$hv_{n-1}$) orientated in clockwise, the aim is to create 'Bridge' edges to combine polygon $P$ and hole $H$ into a new polygon $P_{new}$.

Firstly, create $m \times n$ segments by a pair of vertices. In the pair of vertices, one is a vertex $pv_i$ from PV($pv_0$, $pv_1$,…,$pv_{m-1}$) and the other is a vertex $hv_i$ from HV.

Secondly, compute the length of each segment, and then select the shortest segment as the candidate 'Bridge' edge temporarily. Check whether there have any edge from either the polygon $P$ or the hole $H$ intersects the 'Bridge' edge, if not, the 'Bridge' edge is valid; otherwise, invalid and then test the next shortest segment. These selecting and checking will repeat until a valid 'Bridge' is found.

Finally, combine polygon $P$ and hole $H$ into a new polygon $P_{new}$. Because the two vertices of the 'Bridge' edge are added twice, the new polygon $P_{new}$ has ($m$+$n$+2) vertices.

If there is more than one hole in a polygon $P$, just selecting one of the holes as the hole $H$ and creating a 'Bridge' edge to combine $P$ and $H$ into a new polygon $P_{new}$, and then deem $P_{new}$ as the polygon $P$, also select another holes as the hole $H$. This will be repeated until there are no holes left to obtain the final degenerate polygon. After transforming a polygon with holes to a degenerate polygon, the proposed algorithms will be accepted to generate the final triangulation (Fig.2).

## 3. Applications and Discussion

### 3.1. Tests by Basic Algorithm

In Fig.3(a), a simple polygon with 42 vertices is created based on the Chinese word 'Zhi' which means reaching in English [5], and then it is triangulated by traditional version of ear clipping and the basic algorithm proposed in the paper, as shown in Fig.3(b) and Fig.3(c), respectively. The mentioned traditional version of ear clipping algorithm is the one which locates an ear tip sequentially and does not respect to the special ear tip that has smallest interior angle.

In Table 1, the qualities of triangulations are compared according to the minimum angle of each triangle. The minimum angle of a triangle is between 0° and 60°. And this range from 0° to 60° is divided into four equal intervals, 0°~15°, 15°~30°, 30°~45° and 45°~60°. Obviously, the bigger is the minimum angle in a triangle, the better is the triangle. From the comparison listed in Table 1, the basic algorithm generates better triangulation than that of traditional ear clipping. This conclusion is also true for a multiply-connected polygonal area, Chinese word 'Xi', presented in Fig.4 and also compared in Table 1.

### 3.2. Tests by Improved Algorithm

In Fig.5, a simple polygon with 279 vertices provided by Held [8] is presented, and its triangulations by the improved algorithm with the angle bound set as 0°, 30° and 60°. According to the quality analysis listed in Table 2, when the angle bound is set as 30°, the algorithm with optimization can improve the quality of triangulation a lot comparing to that when the angle bound is 0°. However, the quality improvement due to increase the angle bound from 30° to 60° is not so obvious as that from 0° to 30° . In Fig.6, a polygon with nine holes of 374 vertices provided also by Held [8], is triangulated by the improved algorithm. Same conclusions as that of Fig.6 can also be drawn (Table 2).

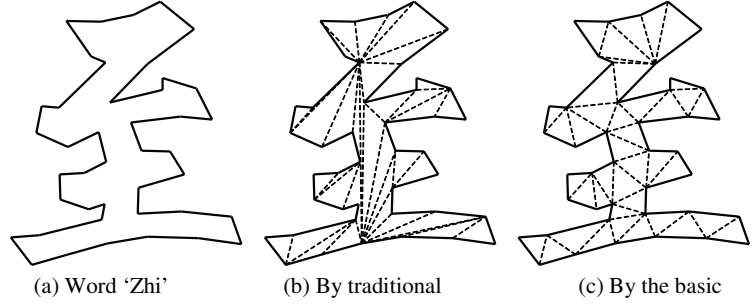

(a) Word 'Zhi'  (b) By traditional  (c) By the basic

**Fig. 3** Chinese word 'Zhi' and its triangulations by traditional and the basic algorithm

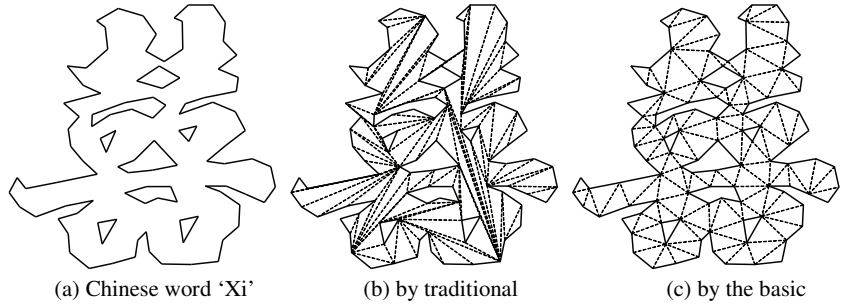

(a) Chinese word 'Xi'  (b) by traditional  (c) by the basic

**Fig. 4** Chinese word 'Xi' and its triangulations by traditional and the basic algorithm

**Table 1** Minimum angle of triangles in triangulations of 'Zhi' and 'Xi'

| Polygon | Algorithm | Minimum angle | | | | |
|---------|-----------|---------------|---|---|---|---|
|         |           | 0°~15° | 15°~30° | 30°~45° | 45°~60° | Average |
| Fig.3   | Traditional | 37.50% | 30.00% | 25.00% | 7.50% | 20.98° |
|         | Basic | 2.50% | 10.00% | 57.50% | 30.00% | 39.26° |
| Fig.4   | Traditional | 55.00% | 30.83% | 11.67% | 2.50% | 16.10° |
|         | Basic | 0.00% | 20.00% | 53.33% | 26.67% | 38.34° |

**Table 2** Minimum angle of triangles in triangulations by the improved algorithm

| Polygon | Algorithm | Minimum angle | | | | |
|---------|-----------|---------------|---|---|---|---|
|         |           | 0°~15° | 15°~30° | 30°~45° | 45°~60° | Average |
| Fig.5   | 0° | 11.91% | 37.18% | 32.85% | 18.05% | 30.04 |
|         | 30° | 5.42% | 37.18% | 37.18% | 20.22% | 32.38 |
|         | 60° | 5.42% | 37.18% | 36.10% | 21.30% | 32.68 |
| Fig.6   | 0° | 15.05% | 34.95% | 37.63% | 12.37% | 29.31 |
|         | 30° | 6.72% | 32.80% | 44.62% | 15.86% | 32.9 |
|         | 60° | 6.72% | 32.80% | 43.82% | 16.67% | 33.17 |

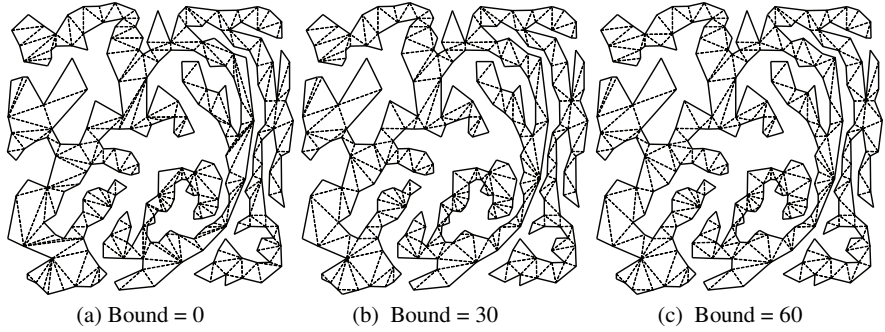

**Fig. 5** Triangulations of simple polygon by the improved algorithm

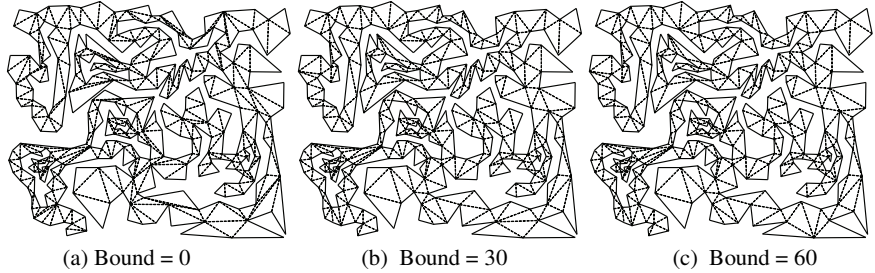

**Fig. 6** Triangulations of polygon with holes by the improved algorithm

## *3.3. Complexity*

In the basic algorithm, the computation of interior angle for all vertices cost $O(n)$ time. And finding all ear tips need $O(n^2)$ time according to Condition 1, although this complexity can be decreased. In the while loop, the most expensive operation is to search the ear tip which has the smallest angle runs in $O(n)$ time. Thus, the while loop spends $O(n^2)$ time on cutting off all ears recursively. Considering this algorithm in whole, it runs in $O(n^2)$ time and $O(n)$ space.

In the improved algorithm, only the procedure of edge swapping is added based on the basic algorithm. So, it is only necessary to analyze the complexity of this part. During edge swapping, finding a suitable neighbor triangle for a newly created triangle costs $O(n)$ time, other operations all run in $O(1)$ time. Hence, only $O(n)$ is spent time on edge swapping and this algorithm with optimization also runs in $O(n^2)$ time and $O(n)$ space.

## 4. Conclusion

Two algorithms based on ear clipping are presented in the paper. In the basic algorithm without optimization, the ear tip with smallest interior angle is always selected and then removed. Better triangulations can be generated by the basic algorithm than traditional ear clipping algorithm which cuts ears off sequentially. To avoid creating sliver triangles in further, edge swapping is adopt based on the proposed basic algorithm. In this algorithm with optimization, an angle bound (recommended as 30) is set to determine whether a newly formed triangle needs swapping edge. The optimization by swapping edges is not implemented after generating whole triangulation but during cutting ears to decrease computations. This algorithm with optimization can reduce sliver triangles effectively.

**Acknowledgments.** This research was supported by the Natural Science Foundation of China (Grant Numbers 40602037 and 40872183). The author would like to thank Prof. M.Held at Universität Salzburg for providing original polygons' data in Fig.5 and Fig.6.